\documentclass[apl,twocolumn,showpacs]{revtex4}
\usepackage{graphicx} \usepackage{bm} \usepackage{amsmath}
\usepackage{amssymb} \newcommand{\etal}{{ \it et al. }}
\newcommand{\D}{{\rm d}}

\newcommand{\I}{{\rm i}}

\begin{document}

\title{Determination of Penetration Depth of Transverse Spin Current in Ferromagnetic Metals by Spin Pumping}

\author{Tomohiro Taniguchi$^{1,2}$, Satoshi Yakata${}^{3}$, Hiroshi
 Imamura$^{2}$, Yasuo Ando${}^{3}$} \affiliation{${}^{1}$Institute
 for Materials Research, Tohoku University, Sendai 980-8577,\\ 
 $^{2}$Nanotechnology Research Institute, National Institute of
 Advanced Industrial Science and Technology, 1-1-1 Umezono, Tsukuba,
 Ibaraki 305-8568, Japan, 
 \\ ${}^{3}$Department of Applied Physics, Graduate School of
 Engineering, Tohoku University, Sendai}

\date{\today} 
\begin{abstract}
 {
   Spin pumping in nonmagnetic/ferromagnetic metal multilayers is studied 
   both theoretically and experimentally. 
   We show that 
   the line widths of the ferromagnetic resonance (FMR) spectrum depend on
   the thickness of the ferromagnetic metal layers, 
   which must not be in resonance 
   with the oscillating magnetic field. 
   We also show that
   the penetration depths of the transverse spin current in ferromagnetic metals can be determined by 
   analyzing the line widths of their FMR spectra. 
   The obtained penetration depths in NiFe, CoFe and CoFeB were 
   3.7 [nm], 2.5 [nm] and 12.0 [nm], respectively.
 }
\end{abstract}

\maketitle


The field of current-driven magnetization dynamics (CDMD) has drawn enormous attention 
because of its potential applications to 
non-volatile magnetic random access memory and microwave devices. 
CDMD is also important from a scientific point of view 
since it provides much information about 
non-equilibrium dynamics of the magnetization 
and 
physics of spin transport 
and 
spin relaxation. 
The concept of CDMD was first proposed by 
Slonczewski \cite{slonczewski96} and independently by Berger \cite{berger96} 
in 1996. 
In the last decade much effort has been devoted 
to studying the physics and applications of CDMD 
both theoretically and experimentally
\cite{sun00,kiselev03}.


One of the most important quantities in CDMD is 
the penetration depth of the transverse spin current $\lambda_{\rm t}$, 
over which 
spin transfer torque is exerted for the magnetization of the free layer. 
However, there is a controversial issue regarding the penetration depth of the transverse spin current. 
One argument is based on the ballistic theory of electron transport, 
and 
its $\lambda_{\rm t}=\pi/|k_{\rm F}^{\uparrow}-k_{\rm F}^{\downarrow}|$,
which is on the order of the lattice constant in conventional ferromagnets 
such as Fe, Co, Ni, and their alloys \cite{stiles02,brataas06}. 
The other argument is based on the Boltzmann theory of electron transport, 
and its $\lambda_{\rm t}$ is on the order of a few nm \cite{zhang02}. 
Urazhdin \etal analyzed the CPP-GMR of noncollinear magnetic multilayers 
using the extended two-series-resistance model 
and concluded that
$\lambda_{\rm t}$=0.8 [nm] for permalloy \cite{urazhdin05}. 
On the other hand, 
Chen \etal analyzed the critical current of the CDMD 
in the Co/Cu/Co trilayer system 
and concluded that 
$\lambda_{\rm t}$=3.0 [nm] for Co \cite{chen06}.

\begin{figure}
\centerline{\includegraphics[width=\columnwidth]{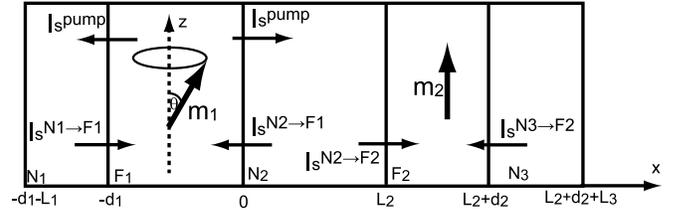}}
 \caption{ Schematic illustration of a nonmagnetic \!\!/\!\! ferromagnetic metal five-layer system. 
    The magnetization of the F${}_{1}$ layer ($\mathbf{m}_{1}$) precesses around the $z$-axis with angle $\theta$. 
    The magnetization of the F${}_{2}$ layer ($\mathbf{m}_{2}$) is fixed with the $z$-axis. 
    The precession of the magnetization in the F${}_{1}$ layer pumps the spin current $\mathbf{I}_{s}^{{\rm pump}}$. 
    The pumped spin current creates spin accumulation in the other layers, 
    and 
    the spin accumulation induces a backflow of spin current $\mathbf{I}_{s}^{{\rm N}\to{\rm F}}$ 
    across each N/F interface.}
\label{fig:model}
\end{figure}

The inverse process of CDMD is spin pumping, 
where spin current is generated 
by precession of magnetization in the ferromagnetic layer 
\cite{tserkovnyak05}. 
Enhancement of the Gilbert damping constant due to spin pumping 
has been extensively studied,
and spin diffusion lengths, 
i.e., 
penetration depths of spin current in nonmagnetic metals, 
have been obtained by 
analyzing the dependence of the enhancement of the Gilbert damping constant 
on the thickness of the nonmagnetic metal layer. 
In spin pumping, 
the direction of the magnetization vector of the pumped spin current is 
perpendicular to the direction of the precessing magnetization vector 
\cite{tserkovnyak05}. 
Let us consider the nonmagnetic/ferromagnetic metal five-layer system 
shown in Fig. \ref{fig:model}. 
Since the magnetization vector of the pumped spin current $\mathbf{I}_{s}^{\rm pump}$ is 
perpendicular to the magnetization vector $\mathbf{m}_{1}$ of the ${\rm F}_{1}$ layer 
and 
the precession angle $\theta$ is very small (about 1 [deg]) in conventional FMR experiments, 
the dominant component of the pumped spin current is 
perpendicular to the magnetization vector $\mathbf{m}_{2}$ of the ${\rm F}_{2}$ layer. 
Therefore, 
it would be possible to determine 
the penetration depth of the transverse spin current in the ${\rm F}_{2}$ layer 
if we could analyze the dependence of the enhancement of the Gilbert damping constant 
on the thickness of the ${\rm F}_{2}$ layer.


In this letter, 
we study spin pumping in
${\rm N}_{1}$/${\rm F}_{1}$/${\rm N}_{2}$/ ${\rm F}_{2}$/${\rm N}_{3}$
five-layer systems shown in Fig. \ref{fig:model} 
both theoretically and experimentally. 
We extend Tserkovnyak's theory of spin pumping
by taking into account the finite penetration depth of the transverse spin current 
and 
show that 
the enhancement of the Gilbert damping constant due to spin pumping 
depends on 
the ratio of the penetration depth $\lambda_{\rm t}$ and the thickness $d_{2}$ of the ${\rm F}_{2}$ layer.  
The motion of the magnetization vector in the ${\rm F}_{2}$ layer is not in resonance with an oscillating magnetic field; 
hence, 
the ${\rm F}_{2}$ layer plays the role of spin absorber.
We also perform FMR experiments in 
Cu/CoFe/Cu/Py/Cu, 
Cu/Py/Cu/CoFe/Cu 
and 
Cu/CoFe/Cu/CoFeB/Cu five-layer systems 
and 
measure line widths $\Delta B$ of energy absorption spectra, 
which are closely related to the Gilbert damping constants. 
The abbreviations CoFe, CoFeB, and Py hereafter refer to
Co${}_{75}$Fe${}_{25}$, 
(Co${}_{50}$Fe${}_{50}$)${}_{80}$B${}_{20}$ 
and Ni${}_{80}$Fe${}_{20}$, respectively. 
Analyzing the dependence of the line width 
on the thickness of the Py, CoFe and CoFeB layers 
that are not in resonance, 
we showed that 
the penetration depths of the transverse spin current 
in the Py, CoFe and CoFeB layers are 
3.7 [nm], 2.5 [nm] and 12.0 [nm], respectively.


Let us begin with an introduction to the theory of spin pumping 
in the nonmagnetic/ferromagnetic metal five-layer system 
shown in Fig. \ref{fig:model} 
with a finite penetration depth of the transverse spin current. 
The pumped spin current generated by precession of the magnetization $\mathbf{m}_{1}$ of 
the ${\rm F}_{1}$ layer is given by
\begin{equation} 
 \mathbf{I}_{s}^{\rm pump}=
 \frac{\hbar}{4\pi}
 \left(
  g_{r({\rm F}_{1})}^{\uparrow\downarrow}
  \mathbf{m}_{1}\times\frac{\D\mathbf{m}_{1}}{\D t}
  +
  g_{i({\rm F}_{1})}^{\uparrow\downarrow}
  \frac{\D\mathbf{m}_{1}}{\D t}
 \right)\ , \label{eq:pump_current}
\end{equation}
where $\hbar$ is the Dirac constant 
and 
$g_{r(i)}^{\uparrow\downarrow}$ is 
the real (imaginary) part of the mixing conductance \cite{tserkovnyak05}. 
The pumped spin current creates spin accumulation in the other layers, 
and the spin accumulation induces a backflow of spin current 
across each N/F interface. 
Although the backflow is obtained from circuit
theory \cite{brataas06,tserkovnyak05}, 
the penetration depth of the transverse spin current $\lambda_{\rm t}$ is assumed to be zero 
in this theory. 
Since we are interested in the effect of the penetration depth of the transverse spin current on spin pumping, 
we explicitly consider the diffusion process of transverse spin accumulation in the ferromagnetic layer. 
The backflow of spin current flowing 
from the ${\rm N}_{i}$ layer to the ${\rm F}_{k}$ layer is expressed as

\begin{equation}
\begin{split}
&\mathbf{I}_{s}^{{\rm N}_{i}\to{\rm F}_{k}}=
 \frac{1}{4\pi}
 \left[
  \frac{2g^{\uparrow\uparrow}g^{\downarrow\downarrow}}{g^{\uparrow\uparrow}+g^{\downarrow\downarrow}}
  \{\mathbf{m}_{k}\cdot
   (\bm{\mu}_{{\rm N}_{i}}-\bm{\mu}_{{\rm F}_{k}}^{{\rm L}})\}
   \mathbf{m}_{k}
\right.
\\ 
&  +
  g_{r({\rm F}_{k})}^{\uparrow\downarrow}\mathbf{m}_{k}\times
   (\bm{\mu}_{{\rm N}_{i}}\times\mathbf{m}_{k})
  +
  g_{i({\rm F}_{k})}^{\uparrow\downarrow}\bm{\mu}_{{\rm N}_{i}}\times\mathbf{m}_{k} 
 \\ 
 &  \left.
  -
  t_{r({\rm F}_{k})}^{\uparrow\downarrow}\mathbf{m}_{k}\times
   (\bm{\mu}_{{\rm F}_{k}}^{{\rm T}}\times\mathbf{m}_{k})
  -
  t_{i({\rm F}_{k})}^{\uparrow\downarrow}\bm{\mu}_{{\rm F}_{k}}^{{\rm T}}\times\mathbf{m}_{k}
 \right]\ , 
               \label{eq:backflow}
\end{split}
\end{equation}
where 
$g^{\uparrow\uparrow(\downarrow\downarrow)}$ is the spin up (down) conductance, 
$t_{r(i)}^{\uparrow\downarrow}$ is the real (imaginary) part of the transmission mixing conductance 
at the F${}_{k}$/N${}_{i}$ interface and 
$\bm{\mu}_{{\rm N}_{i}}$ is the spin accumulation in the ${\rm N}_{i}$ layer \cite{brataas06}. 
The longitudinal spin accumulation in the ${\rm F}_{k}$ layer is denoted by $\bm{\mu}_{{\rm F}_{k}}^{{\rm L}}$. 
The last two terms express contributions from transverse spin accumulation $\bm{\mu}_{{\rm F}_{k}}^{{\rm T}}$ 
in the ${\rm F}_{k}$ layer.
 

The spin accumulation in a ferromagnetic layer is defined by 
the non-equilibrium distribution matrix at a given energy $\varepsilon$, 
$\hat{f}(\varepsilon)=f_{0}\hat{1}+\bm{f}\cdot\hat{\bm{\sigma}}$ \cite{brataas06}, 
where $\bm{f}=f_{x}\bm{{\rm t}}_{1}+f_{y}\bm{{\rm t}}_{2}+f_{z}\mathbf{m}$. 
Here we introduce the orthogonal unit vectors in spin space 
($\bm{{\rm t}}_{1},\bm{{\rm t}}_{2},\mathbf{m}$). 
The non-equilibrium charge distribution is represented by $f_{0}=(f^{\uparrow}+f^{\downarrow})/2$. 
On the other hand, 
$f_{z}=(f^{\uparrow}-f^{\downarrow})/2$ is the difference in non-equilibrium distribution 
between spin-up and spin-down electrons, 
and 
$f_{x}$ and $f_{y}$ are the non-equilibrium distributions of the transverse spin components. 
The spin accumulation is defined as 
$\bm{\mu} = \int_{\varepsilon_{\rm F}} \D\varepsilon {\rm Tr}[\hat{\bm{\sigma}}\hat{f}]$ \cite{tserkovnyak05}. 
The spin accumulation in the nonmagnetic layer is defined in a similar way. 


The spin accumulation in a nonmagnetic layer, $\bm{\mu}_{\rm N}$, 
obeys the diffusion equation \cite{valet93}, 
and is expressed as a linear combination of 
$\exp(\pm x/\lambda_{\rm sd(N)})$, 
where $\lambda_{\rm sd(N)}$ is the spin diffusion length.
The spin current in a nonmagnetic layer is given by 
\begin{equation}
  \mathbf{I}_{s}^{{\rm N}}
  =
  -\frac{\partial}{\partial x}
  \frac{\hbar S\sigma_{{\rm N}}}{2e^{2}}
  \bm{\mu}_{{\rm N}}\ ,
\end{equation}
where $S$ is the cross section area of the system, 
$\sigma_{\rm N}$ is the conductivity 
and $e$ is the absolute value of the electron charge. 
The spin current in the ${\rm N}_{3}$ layer is equal to 
$-\mathbf{I}_{s}^{{\rm N}_{3}\to{\rm F}_{2}}$ at $x=L_{2}+d_{2}$ 
because of the continuity of the spin current, 
and 
vanishes at $x=L_{2}+d_{2}+L_{3}$ (see Fig. \ref{fig:model}). 
Using above boundary conditions, 
we obtain the spin accumulation in the N${}_{3}$ layer. 


The longitudinal spin current in a ferromagnetic layer is given by
\begin{equation}
  (\mathbf{m}\cdot\mathbf{I}_{s}^{{\rm F}})\mathbf{m}
  =
  -\frac{\partial}{\partial x}
  \frac{\hbar S}{2e^{2}}
  (
    \sigma^{\uparrow}\mu_{{\rm F}}^{\uparrow}
    -
    \sigma^{\downarrow}\mu_{{\rm F}}^{\downarrow}
   )\mathbf{m}\ ,
 \label{eq:diff_ferrol}
\end{equation}
where 
$\mu_{{\rm F}}^{\uparrow(\downarrow)}=\int_{\varepsilon_{{\rm F}}}\D\varepsilon f^{\uparrow(\downarrow)}$ is 
the electro-chemical potential for the spin-up (spin-down) electrons 
and $\sigma^{\uparrow(\downarrow)}$ is the conductivity of spin-up (spin-down) electrons. 
The polarization of spin-dependent conductivity is defined as 
$\beta=(\sigma^{\uparrow}-\sigma^{\downarrow})/(\sigma^{\uparrow}+\sigma^{\downarrow})$. 
The longitudinal spin current in the ${\rm F}_{2}$ layer is equal to 
$\mathbf{m}_{2}\cdot\mathbf{I}_{s}^{{\rm N}_{2}\to{\rm F}_{2}}$ at $x=L_{2}$ 
and 
$-\mathbf{m}_{2}\cdot\mathbf{I}_{s}^{{\rm N}_{3}\to{\rm F}_{2}}$ at $x=L_{2}+d_{2}$ 
because of the continuity of the spin current. 
Solving the diffusion equation \cite{valet93} with the above boundary conditions, 
we obtain longitudinal spin accumulation in the F${}_{2}$ layer. 
The longitudinal spin accumulation is expressed as a linear combination of 
$\exp(\pm x/\lambda_{\rm sd(F_{L})})$, 
where $\lambda_{\rm sd(F_{L})}$ is the longitudinal spin diffusion length. 


We assume that the transverse spin accumulation obeys the following equation \cite{zhang02}:
\begin{equation}
  \frac{\partial^{2}}{\partial x^{2}}\bm{\mu}_{{\rm F}}^{{\rm T}}
  =
  \frac{1}{\lambda_{J}^{2}}\bm{\mu}_{{\rm F}}^{{\rm T}}\times\mathbf{m} 
  + 
  \frac{1}{\lambda_{{\rm sd(F_{T})}}^{2}}\bm{\mu}_{{\rm F}}^{{\rm T}}\ ,
  \label{eq:diff_ferrot}
\end{equation}
where $\lambda_{J}=\sqrt{(D^{\uparrow}+D^{\downarrow})\hbar/(2J)}$ 
and $\lambda_{\rm {sd(F_{T})}}$ is the transverse spin diffusion length. 
Here $J$ represents the strength of the exchange field. 
The transverse spin accumulation is expressed as a linear combination of 
$\exp(\pm x/l_{+})$ and $\exp(\pm x/l_{-})$, 
where $1/l_{\pm}=\sqrt{(1/\lambda_{{\rm sd(F^{T})}}^{2})\mp(\I/\lambda_{J}^{2})}$. 
Therefore, 
we define the penetration depth of the transverse spin current $\lambda_{\rm t}$ by 
\begin{equation}
  \frac{1}{\lambda_{\rm t}}
  =
  {\rm Re}\left[\frac{1}{l_{+}}\right]. 
\end{equation}
The transverse spin current in a ferromagnetic layer is expressed as
\begin{equation}
\mathbf{m}\times(\mathbf{I}_{s}^{{\rm F}}\times\mathbf{m})=
 -\frac{\partial}{\partial x}\frac{\hbar S\sigma^{\uparrow\downarrow}}{2e^{2}}
 \bm{\mu}_{{\rm F}}^{{\rm T}}\ ,
\end{equation}
where 
$\sigma^{\uparrow\downarrow}
 =(1/2)(\sigma^{\uparrow}/(1+\beta^{\prime})+\sigma^{\downarrow}/(1-\beta^{\prime}))$. 
Here
$\beta^{\prime}=(D^{\uparrow}-D^{\downarrow})/(D^{\uparrow}+D^{\downarrow})$ is 
the polarization of the spin-dependent diffusion constants, 
$D^{\uparrow}$ and $D^{\downarrow}$ \cite{zhang02}. 
For simiplicty, 
we assume that 
$\beta=\beta^{'}$. 
The transverse spin accumulation in a ferromagnetic layer is obtained by 
solving Eq. \eqref{eq:diff_ferrot} with boundary conditions 
satisfying the continuity of the spin current at the N/F interface.


Solving the diffusion equations of the spin accumulations 
of the N${}_{3}$ and F${}_{2}$ layers, 
the backflow at the N${}_{2}$/F${}_{2}$ interface 
is re-written as 
\begin{equation}
\begin{split}
  \mathbf{I}_{s}^{{\rm N}_{2}\to{\rm F}_{2}}
   = 
  \frac{1}{4\pi} &
  \left[
    g_{({\rm F}_{2})}^{*}(\mathbf{m}_{2}\cdot\bm{\mu}_{{\rm N}_{2}})\mathbf{m}_{2}
  \right.
\\
    &+   \left.
    \tilde{g}_{r({\rm F}_{2})}^{\uparrow\downarrow}\mathbf{m}_{2}\times(\bm{\mu}_{{\rm N}_{2}}\times\mathbf{m}_{2})
    +
    \tilde{g}_{i({\rm F}_{2})}^{\uparrow\downarrow}\bm{\mu}_{{\rm N}_{2}}\times\mathbf{m}_{2}
  \right]\ ,
\end{split}
\end{equation}
where the conductance $g_{({\rm F}_{2})}^{*}$ is given in Ref. \cite{tserkovnyak05}, 
and depends on the ratio $d_{2}/\lambda_{\rm sd(F_{L})}$. 
Similarly, 
the renormalized mixing conductances, 
$\tilde{g}_{r,i({\rm F}_{2})}^{\uparrow\downarrow}$, 
depend on the ratio $d_{2}/l_{+({\rm F}_{2})}$. 
If the thickness of the N${}_{3}$ layer is thin enough compared to its spin diffusion length, 
$\tilde{g}_{r,i({\rm F}_{2})}^{\uparrow\downarrow}$ are given by 
\begin{equation}
  \begin{pmatrix}
    \tilde{g}_{r({\rm F}_{2})}^{\uparrow\downarrow} \\
    \tilde{g}_{i({\rm F}_{2})}^{\uparrow\downarrow}
  \end{pmatrix}
  =
  \frac{1}{\Delta}
  \begin{pmatrix}
    K_{1} & K_{2} \\
    -K_{2} & K_{1}
  \end{pmatrix}
  \begin{pmatrix}
    g_{r({\rm F}_{2})}^{\uparrow\downarrow} \\
    g_{i({\rm F}_{2})}^{\uparrow\downarrow}
  \end{pmatrix}\ ,
\end{equation}
where $\Delta=K_{1}^{\ 2}+K_{2}^{\ 2}$ and 
$K_{1}$ and $K_{2}$ are given 
\begin{equation}
\begin{split}
  K_{1}
  =
  1 
  & + 
  t_{r({\rm F}_{2})}^{\uparrow\downarrow}
  {\rm Re}\left[\frac{1}{g_{\rm t}\tanh(d_{2}/l_{+})}\right]
\\
  &+
  t_{i({\rm F}_{2})}^{\uparrow\downarrow}
  {\rm Im}\left[\frac{1}{g_{\rm t}\tanh(d_{2}/l_{+})}\right]\ ,
\end{split}
\end{equation}
\begin{equation}
  K_{2}
  =
  t_{i({\rm F}_{2})}^{\uparrow\downarrow}
  {\rm Re}\left[\frac{1}{g_{\rm t}\tanh(d_{2}/l_{+})}\right]
  -
  t_{r({\rm F}_{2})}^{\uparrow\downarrow}
  {\rm Im}\left[\frac{1}{g_{\rm t}\tanh(d_{2}/l_{+})}\right]\ ,
\end{equation}
where $g_{\rm t}/S=h/2e^{2}\rho_{\rm F_{2}}l_{+}$ and 
$\rho_{\rm F_{2}}$ is the resistivity of the F${}_{2}$ layer. 
The mixing conductance of the F${}_{1}$ layer 
in Eqs. (\ref{eq:pump_current}) and (\ref{eq:backflow}) is 
also replaced by the renormalized conductance.

\begin{figure}
  \centerline{\includegraphics[width=0.6\columnwidth]{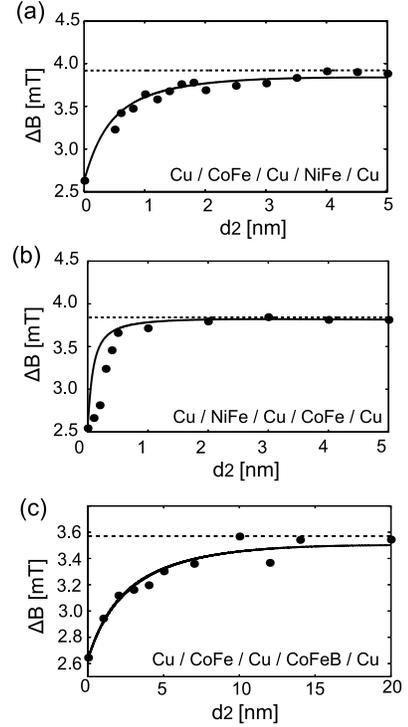}}
  \caption{
    The dependences of the line width of the FMR power absorption spectra, $\Delta B$, 
    on the thickness of the F${}_{2}$ layer, $d_{2}$.
    Materials of the F${}_{2}$ layer are 
    (a) Ni${}_{80}$Fe${}_{20}$, 
    (b) Co${}_{75}$Fe${}_{25}$ 
    and 
    (c) (Co${}_{50}$F${}_{50}$)${}_{80}$B${}_{20}$, respectively. 
    The filled circles represent experimental data 
    and 
    the solid lines are fit to the experimental data 
    according to the theory with the finite penetration depth of the transverse spin current 
    $\lambda_{\rm t}$. 
    The dotted lines represent the case of $\lambda_{\rm t}=0$. 
 }
 \label{fig:fmr}
\end{figure}


We assume that 
spin-flip scattering in the N${}_{2}$ layer is so weak 
that we can neglect the spatial variation of the spin current 
in the N${}_{2}$ layer. 
Then we have 
$\mathbf{I}_{s}^{\rm pump}-\mathbf{I}_{s}^{{\rm N}_{1}\to{\rm F}_{1}}=\mathbf{I}_{s}^{{\rm N}_{2}\to{\rm F}_{2}}$, 
and the spin accumulation in the N${}_{2}$ layer can be determined \cite{tserkovnyak05}. 
The torque acting on the magnetization of the F${}_{1}$ layer 
is given by 
$\mathbf{m}_{1}\times\{(\mathbf{I}_{s}^{\rm pump}-\mathbf{I}_{s}^{{\rm N}_{2}\to{\rm F}_{1}})\times\mathbf{m}_{1}\}$, 
which yields the following modified Landau-Lifshitz-Gilbert (LLG) equation \cite{tserkovnyak05,taniguchi07} :
\begin{equation}
  \frac{\D\mathbf{m}_{1}}{\D t}
  =
  -\gamma_{\rm eff}\mathbf{m}_{1}\times\bm{{\rm B}}_{{\rm eff}}
  +
  \frac{\gamma_{\rm eff}}{\gamma}
  (\alpha_{0}+\alpha^{\prime})
  \mathbf{m}_{1}\times\frac{\D\mathbf{m}_{1}}{\D t}\ ,
  \label{eq:llg}
\end{equation}
where $\bm{{\rm B}}_{{\rm eff}}$ is the effective magnetic field, 
$\gamma$ is the gyromagnetic ratio, 
$\alpha_{0}$ is the Gilbert damping constant intrinsic to the ferromagnetic metal, 
and $\alpha^{\prime}$ is the enhancement of the Gilbert damping constant due to spin pumping. 
The Gilbert damping constant is related to the line width of the FMR absorption spectrum via 
\cite{vonsovskii}
\begin{equation}
  \Delta B 
  = 
  \Delta B_{0} + \frac{2\omega}{\sqrt{3}\gamma}\alpha^{\prime}\ ,
\end{equation}
where $\omega=2\pi f$ is the angular velocity of the oscillating magnetic field. 
We notice that 
the effects of the N${}_{1}$ and N${}_{3}$ layers are quite small 
because, as mentioned below, 
the thickness of these layers are thin enough 
compared to its spin diffusion length 
in our experiments. 
Assuming that $g_{r}^{\uparrow\downarrow}\gg g_{i}^{\uparrow\downarrow}$ \cite{tserkovnyak05}, 
in the limit of $\theta\to0$, we find 
\begin{equation}
  \Delta B - \Delta B_{0}
  \simeq
  \frac{\hbar\omega}{2\sqrt{3}\pi Md_{1}S}
  \frac{\tilde{g}_{r({\rm F}_{1})}^{\uparrow\downarrow}\tilde{g}_{r({\rm F}_{2})}^{\uparrow\downarrow}}
    {\tilde{g}_{r({\rm F}_{1})}^{\uparrow\downarrow}+\tilde{g}_{r({\rm F}_{2})}^{\uparrow\downarrow}}\ ,
  \label{eq:line_width}
\end{equation}
where $\tilde{g}_{r({\rm F}_{i})}^{\uparrow\downarrow}(i=1,2)$ is 
the real part of the renormalized mixing conductance of the $i$-th ferromagnetic layer. 
We should note that 
if we neglect the transverse spin accumulation in the ferromagnetic layer 
the mixing conductances are not renormalized, 
and that the line width $\Delta B$ does not depend on 
the thickness of the F${}_{2}$ layer \cite{tserkovnyak05}. 
This is due to the fact that 
the dominant component of the pumped spin current is 
perpendicular to the magnetization vector $\mathbf{m}_{2}$ of the ${\rm F}_{2}$ layer 
in our experiment.


We performed FMR experiments on the three different 
${\rm N}_{1}$/${\rm F}_{1}$/ ${\rm N}_{2}$/${\rm F}_{2}$/${\rm N}_{3}$
five-layer systems shown in Fig. \ref{fig:model} 
\cite{mizukami01a}. 
Nonmagnetic layers are made of Cu. 
The combinations of the ferromagnetic layers
(F${}_{1}$,F${}_{2}$) of each system are 
(a) (CoFe,Py), 
(b) (Py,CoFe) and (c) (CoFe,CoFeB). 
The samples were deposited on Corning 1737 glass substrates 
using an rf magnetron sputtering system in an ultrahigh vacuum 
below 4$\times$10${}^{-6}$ [Pa] and 
cut to 5 [mm${}^{2}$]. 
The Ar pressure during deposition was 0.077 [Pa]. 
The thickness of all Cu layers are 5 [nm]. 
The thickness of F${}_{1}$ layers is 5 [nm] for sample (a) and (b), 
and 10 [nm] for sample (c).  
The FMR measurements were carried out 
using an X-band microwave source ($f=9.4$[GHz]) at room temperature. 
The microwave power, modulation frequency, and modulation field are 
1 [mW], 10 [kHz], and 0.1 [mT], respectively. 
The precession angles of all samples are estimated to be 1 [deg]. 
The resistivity $\rho_{\rm F}$ of Py, CoFe and CoFeB are 
241 [$\Omega$nm], 94 [$\Omega$nm] and 1252 [$\Omega$nm], respectively. 
The magnetizations ($4\pi M$) of Py and CoFe are 0.76 [T] and 2.1 [T], respectively. 
The gyromagnetic ratio is $1.8467\times 10^{11}$ [Hz/T] for all systems


In Figs. \ref{fig:fmr} (a), (b) and (c), 
the measured line widths of the FMR absorption spectra $\Delta B$ are plotted 
with filled circles 
against the thickness of the ${\rm F}_{2}$ layer, $d_{2}$. 
The solid lines are fit to the experimental data 
according to the theory with the finite penetration depth of the transverse spin current $\lambda_{\rm t}$. 
The dotted lines represent the calculated $\Delta B$ 
in the case of $\lambda_{\rm t}=0$ \cite{tserkovnyak05}. 


Parameters other than $\lambda_{\rm t}$ are determined as follows. 
The mixing conductances per unit area of the combinations 
($g_{r({\rm F}_{1})}^{\uparrow\downarrow}/S$, $g_{r({\rm  F}_{2})}^{\uparrow\downarrow}/S$) 
are assumed to be 
(a) (48.0, 38.0), (b) (15.2, 17.0) and (c) (48.0, 128.0) [nm${}^{-2}$]. 
Although these values are determined by fitting, 
they have good agreement with the \textit{ab\ initio} caluclations \cite{brataas06}. 
For simplicity, 
we assume that 
$t_{r}^{\uparrow\downarrow}=t_{i}^{\uparrow\downarrow}$
where values of $t_{r,i}^{\uparrow\downarrow}/S$ of Py, CoFe and CoFeB are taken to be 
4.0 [nm${}^{-2}$], 6.0 [nm${}^{-2}$] and 0.8 [nm${}^{-2}$], respectively. 
The longitudinal spin diffusion lengths are 
5.5 [nm] for Py and 12 [nm] for CoFe and CoFeB, respectively \cite{reilly99,fert99}. 
The polarizations of conductance $\beta$ are 
0.73 for Py, 
0.65 for CoFe 
and 
0.56 for CoFeB, respectively \cite{reilly99,fert99,oshima02}. 
The transverse spin diffusion lengths are given by 
$\lambda_{\rm sd(F_{T})}=\lambda_{\rm sd(F_{L})}/\sqrt{1-\beta^{2}}$ 
\cite{zhang02}. 
We take 
$g_{i}^{\uparrow\downarrow}/S=1.0$ [nm${}^{-2}$],
$2g^{\uparrow\uparrow}g^{\downarrow\downarrow}/(g^{\uparrow\uparrow}+g^{\downarrow\downarrow})S=20.0$ [nm${}^{-2}$]
\cite{tserkovnyak05,brataas06} for all systems; 
these are not important parameters for fitting the experimental results. 
The spin diffusion length and resistivity of Cu are taken to be 
500 [nm] and 21 [$\Omega$nm] \cite{bass07}. 
%
%
The obtained values of $\lambda_{\rm t}$ are 
3.7 [nm] for Py, 
2.5 [nm] for CoFe and 
12.0 [nm] for CoFeB, respectively. 
Our results agree quite well with the prediction 
based on the Boltzmann theory of electron transport
\cite{zhang02}. 


In conclusion, 
we analyzed spin pumping in 
Cu/CoFe/Cu/Py/Cu,
Cu/Py/Cu/CoFe/Cu and 
Cu/CoFe/Cu/CoFeB/Cu five-layer systems 
both theoretically and experimentally. 
We showed that 
the enhancement of the Gilbert dumping constant due to spin pumping depends on 
the ratio of the penetration depth $\lambda_{\rm t}$ 
and 
the thickness of the ferromagnetic layers, 
which is not in resonance with the oscillating magnetic field. 
We measured the line widths of FMR absorption spectra, 
which are closely related to the Gilbert dumping constant. 
Analyzing the experimental results, 
we showed that 
the penetration depths of the transverse spin current 
in Py, CoFe and CoFeB are 
3.7 [nm], 2.5 [nm] and 12.0 [nm], respectively. 
Our results support the Boltzmann theory of transverse spin current
\cite{zhang02}.


The authors would like to acknowledge the valuable discussions they had with P. M. Levy. 
This work was supported by CREST and NEDO.


\end{document}